\documentclass[journal]{IEEEtran}
\usepackage{graphicx}
\usepackage{pdfpages}

\ifCLASSINFOpdf
\else
\fi
\begin{document}
%
\title{A Literature Survey on Ontology of Different Computing Platforms in Smart Environments}
%

\author{Souvik~Sengupta,~\textit{PhD student, UPC BarcelonaTech,}
        Jordi~Garcia,~\textit{Associate Professor, UPC BarcelonaTech,}
        and~Xavi~Masip-Bruin~\textit{Associate Professor, UPC BarcelonaTech.}
\thanks{S.Sengupta, J.Garcia, and X.Masip-Bruin are with the Advanced Network Architectures Lab (CRAAX), the Department of Computer Architecture (Departament d'Arquitectura de Computadors, DAC) at the {\em Universitat Politecnica de Catalunya}, UPC BarcelonaTech. email: [souvik, jordig, xmasip]@ac.upc.edu.}}
\maketitle

\begin{abstract}
\textit{Smart environments integrates various types of technologies, including cloud computing, fog computing, and the IoT paradigm. In such environments, it is essential to organize and manage efficiently the broad and complex set of heterogeneous resources. For this reason, resources classification and categorization becomes a vital issue in the control system. In this paper we make an exhaustive literature survey about the various computing systems and architectures which defines any type of ontology in the context of smart environments, considering both, authors that explicitly propose resources categorization and authors that implicitly propose some resources classification as part of their system architecture. 
As part of this research survey, we have built a table that summarizes all research works considered, and which provides a compact and graphical snapshot of the current classification trends.
The goal and primary motivation of this literature survey has been to understand the current state of the art and identify the gaps between the different computing paradigms involved in smart environment scenarios. As a result, we have found that it is essential to consider together several computing paradigms and technologies, and that there is not, yet, any research work that integrates a merged resources classification, taxonomy or ontology required in such heterogeneous scenarios.}
\end{abstract}


%
\IEEEpeerreviewmaketitle
%
%
%
%

\section{Introduction}
\label{sec:intro}
Information and communication technologies are experiencing fast and relevant changes in our modern society. Network technologies are becoming ubiquitous and more efficient, and computing services are delivering higher performance as well as faster response times, rising the development of new and more sophisticated computing paradigms. Such progression in the advancement of computing and communication technologies is pushing the emergence of the smart computing era. We are already surrounded by several smart concepts, such as smart cities, smart industry, smart farms, all in all, building the concept of smart environments. In such contexts, different types of computing devices are integrated, together with multiple types of sensors and actuators. Sensors networks are continuously capturing and producing large volumes of structured and unstructured data, thus including big data technologies into the smart environments.\\
Indeed, smart environments have been defined as {\em a physical world that is richly and invisibly interwoven with sensors, actuators, displays, and computational elements, embedded seamlessly in the everyday objects of our lives, and connected through a continuous network} \cite{1}, or also as {\em a small world where different kinds of smart devices are continuously working to make inhabitants' lives more comfortable} \cite{2}. The concept of smart computing comes from the ubiquitous computing, whereby integrating various kind of existing technologies (such as cloud computing, fog computing, the Internet of Things (IoT), or Wireless Sensor Networks (WSN)) builds up the smart environment paradigm.\\
So defining efficient smart environments essentially means appropriately integrating various kind of existing technologies, where some of such technologies (for instance, cloud computing) provide the best facilities to store and process big data, and some other (for example, the IoT or the WSN) provide the best facilities to producing and capturing those data from various kind of sources of the real world. Furthermore, additional technologies (for instance, fog computing or edge computing) provide the most appropriate facilities to process those data near to the end users or edge devices, speeding up performance as well as reducing network traffic through the system. Considering and combining all these facilities provided by the various existing technologies, helps creating better and more efficient smart environment systems.\\
In order to run such complex systems efficiently, it is essential to organize and manage the broad set of heterogeneous resources properly. For this reason, resources organization becomes a vital issue in the control system. Efficient resources management and organization requires deep knowledge about the type and features of those system resources, and proper categorization. Our primary intention in this literature survey is to identify the resources of the different computing paradigms that build up smart environments, and also to find out the features and behaviors of those resources. We give our highest concern to finding the possible resources classifications for those computing paradigms, as well as looking for the resource and system ontologies that have been defined by the various researchers.\\
In this research work, we have focused on those computing platforms and system resources related to smart environments and tried to identify the technological gaps and differences between resources of the various computing platforms. We have surveyed those research works which define somehow the resources of different computing paradigms and which present some sort of classification, taxonomy or ontology. As a result of our survey, we have found that, for smart environments, it is essential to consider together several computing paradigms and technologies, and that there is not, yet, any research work that integrates a merged resources classification, taxonomy or ontology required in such heterogeneous smart scenarios.\\
The rest of this paper is organized as follows. Section \ref{sec:sos} positions the scope of this literature survey by revisiting the different definitions of {\em classification}, {\em categorization}, {\em taxonomy}, and {\em ontology}, and describing the considered scenarios of analysis. In Section \ref{sec:liter}, we present the extensive literature survey organized according to the various computing paradigm considered: fog computing, IoT, and cloud computing. By following up the list of proposals, in Section \ref{sec:ana} we present our analysis of the different research works and summarize them in a table. Some interesting findings are discussed in Section \ref{sec:synop} and, finally, some concluding remarks are presented in Section \ref{sec:conc}.

\section{Scope of the Survey}
\label{sec:sos}
Different research works have been published describing systems in terms of \textit{classification}, \textit{categorization}, \textit{taxonomy}, or even \textit{ontology}. Before diving into the details of such literature, in this section, we will try to understand and clarify this terminology and propose our interpretation of these terms. And then, we will describe the scenarios of analysis that will determine the scope of our literature survey.

\subsection{Classification, Categorization, Taxonomy, and Ontology}
\label{sec:defi}
Knowing the system characteristics, features and behaviors, as well as other related vital aspects, is essential for a proper system understanding. In our literature survey, we have observed that several researchers present some taxonomy by considering a categorization or classification of their system resources and services, or even propose some ontology for their systems. In \cite{3}, the authors define \textit{classification} as the set of concepts which have been considered for grouping entities according to their similarity. They also said that, the taxonomy contains a "subsumption hierarchy" in the form of transitive \textit{subclassOf} relations, i.e., each object of a class can be assumed to be also an object of all parent categories. Whereas in \cite{4}, authors define \textit{taxonomy} as \textit{the science of categorization, or classification, of things based on a predefined system; typically, taxonomy contains a controlled vocabulary with a hierarchical tree-like structure}. So according to them the taxonomy is a path to build a systematic model, by organizing the features into categories and subcategories. They also said that, according to a philosophical view, an \textit{ontology} is the study of the kind of things that exist in a real scenario, which defines the type of things and their relations. In \cite{5}, the authors say that ontology has mostly come to mean one of the two related things: {\em i)} it is a representation vocabulary, often specialized in some domain or subject matter, or {\em ii)} it is used to refer to a body of a knowledge for describing some domain. In \cite{6}, the authors say that \textit{by implicitly or explicitly, every knowledge model is committed to some conceptualization, and an ontology is the explicit specifications of this conceptualization}. So an ontology is the collection of various kinds of entities and relations among them.

From the above definitions, we conclude that \textit{classification} organizes concepts in order to build the \textit{taxonomy} \cite{7}, what means that \textit{classification} is the subset of \textit{taxonomy}. Whereas the term \textit{ontology} is much broader and larger in scope \cite{8}. For instance, as an illustrative example, in an ontological forest, a taxonomy is a tree. Or, regarding graph theory, we can represent a taxonomy as the tree and an ontology as the graph, where all the edges are to be considered as the relations between different vertices. In this paper, our focus is to identify how researchers propose their ontologies, by following the various classifications, categorizations and taxonomies of different computing paradigms.

\subsection{Considered Research Scenarios}
\label{subsec:crs}
Cloud computing initiated a new era of computation, where most of the work (i.e. computation, data processing, etc.) is performed at the cloud data centers, which are often geographically centralized. Besides, vast numbers of IoT devices and sensors are increasingly being deployed everywhere. Almost all modern devices are connected to the Internet, and many of them have additional sensors and actuators. In fact, one of the leading research and advisory company, Gartner Inc. \cite{9}, forecasts that by 2020 almost 20.4 billion connected things will be in use worldwide. Also by 2022, M2M traffic flows are expected to constitute up to 45\% of the whole Internet traffic \cite{10}. Beyond these predictions, McKinsey Global Institute reported in 2015 that the number of connected machines had grown 300\% over the last five years. Traffic monitoring of a cellular network in the US also showed an increase of 250\% for M2M traffic volume in 2011 \cite{11}. Also, according to the Cisco reports, they predict that 50 billion objects and devices will be connected to the Internet by 2020. However, more than 99 percent of today's available things in the world remain unconnected \cite{12}. According to a Navigant Research report, the number of installed smart meters around the world will grow to 1.1 billion by 2022 \cite{13}. Another report from Automotive News states that the number of cars connected to the Internet worldwide will increase from 23 million in 2013 to 152 million in 2020 \cite{14}. So, it is clear that IoT is going to rule over the modern technology, but most importantly, these IoT devices and sensors are highly distributed over the network, along with the agile and latency-sensitive service requirements. So, for the cloud data centers, in case of latency-sensitive service requirements, it is quite apparent that they might fail to deal with high processing and storage demand for this vast set of IoT devices at the edge. As a result, congested network, high latency in service delivery, and poor Quality of Service (QoS) are being experienced \cite{15}. To overcome these problems, researchers come with a new computing paradigm, fog computing, which consists of adding some processing capabilities between the cloud data center and the IoT devices/sensors. Fog computing is one emerging technology, so there are still lots of work to do in this domain. Still, many research work is going on to make some transparent and standard architectures for this platform, which will allow being quickly adopted by any use-case. The primary objective of fog computing is to extend the cloud computing facilities to the edge of the network \cite{15,16}, turning fog computing to be agiler. But, interestingly, most of the research works justify that fog computing is not going to rule over the cloud computing, rather than together, they are going to solve many problems and will provide better facilities to the next computing and networking platforms \cite{16}. In addition to the cloud, fog, and IoT paradigms, which define the core of our interests, some other related technologies will complete the whole picture, such as grid computing, mobile edge computing, and wireless sensor networks, which will also be considered in this survey.

\section{Literature Survey}
\label{sec:liter}
In this section, we briefly describe the different research works which have considered somehow the resources categorization, classification, taxonomy, ontology, or just some sort of resources management, all of them in the scope of smart environments. Some authors explicitly use any of these terms in their works, but some others propose implicitly some resources classification as part of their system architecture organization. We have tried to include all of them in this literature survey. In section \ref{subsec:fcp} we address those research works in the area of fog computing. In section \ref{subsec:iotp} we discuss the IoT related research works, in section \ref{subsec:ccp} we focus on the cloud computing related literature and, in section \ref{subsec:ocp}, we consider those additional research works on the areas of grid computing, WSN, mobile edge computing, or other related technologies.

\subsection{Fog Computing Platforms}
\label{subsec:fcp}
Several authors have published their work with some contributions related to resources categorization in the scope of fog computing. In \cite{16,17,18,19} the authors present mainly some architectural contributions and, as part of these, they address some classification issues. In \cite{15,20,21} the researchers propose explicitly one taxonomy for fog computing platforms. In addition, we have observed that many researchers focus on identifying features, behaviors and attributes in fog computing, as in \cite{18,19,20,22,23,24}, where they discuss their challenges and characteristics. Interestingly, in \cite{21} they present a taxonomy for a connected object, which is mostly suggested as the taxonomy for edge or fog devices. In the next few subsections, we discuss individually those research works which have been done around the fog computing paradigm.\\

\subsubsection{Bonomi et al.}
Bonomi et al. have made important contributions in the areas of fog computing and, consequently, they have several works presenting different aspects about their architecture. In this literature survey we are considering two of their works. In \cite{16}, the authors present fog computing as a distributed architecture, where fog relies on technological components for scalable virtualization of the essential resource classes, such as computing, storage, and networking. By following up two use-cases, they propose the software architecture for a fog computing platform. The main components of their proposed architecture are various physical resources (computation resources, network resources, and storage resources), the abstraction layer, the fog service orchestration layer, and some APIs for applications. They propose the abstraction layer to provide a uniform and programmable interface by hiding the platform heterogeneity for monitoring and managing, not only the resources, but also the various hypervisors, OSes, and service containers. By adopting different techniques to support virtualization in a fog computing architecture, this layer allows running OSes or service containers on a physical machine to improve resource utilization. In this paper, authors have not explicitly defined any classification or taxonomy of the system resources, but reviewing their architecture, we can get an idea about the classification of physical resources as well as identify the attributes, features and behaviors of the fog resources. Whereas in \cite{24}, the authors initially discuss on various use-cases and real scenarios, where fog is playing a predominant role. Considering all these scenarios, the authors identified the characteristics of fog computing paradigm. They said that fog computing has some key attributes. As fog brings to the cloud facilities to the edge of the network, it is quite relevant that fog is providing the services to the end users, within low latency or delay.  Also, it has some unique features like location awareness, heterogeneity, support to virtualization, mobility, and so on. In this paper, the authors add that fog computing enables the facility of interoperability and federation across the various service domains and facilitates the on-line analytics and interplay by processing the data close to the source. They discuss the primary attributes and features in the fog computing paradigm and characterize the participants of fog computing, such as providers and users. Focusing on their work, we can also identify the characteristics, attributes and behaviors of their system resources. Though they have not explicitly mentioned the classification, taxonomy or ontology of their system resources, by discussing their system's resources characteristics, attributes, behaviors, and resource owner’s role, it helps us identifying their system resource classification.

\subsubsection{Munir et al.}
In this paper \cite{17}, the authors present a novel integrated fog-cloud architecture by considering the benefits of IoT, fog, and cloud computing platforms. Their proposed architecture is an energy efficient reconfigurable layered architecture and, according to them, this architecture can easily fit with various kind of fog computing applications. The layered architecture facilitates abstraction. In their architecture, the bottom section contains two components: the hardware layer and the reconfiguration layer. The virtualization layer resides on top of these two layers, and it abstracts the hardware resources of the edge server from the higher application service layer, where the virtual machines run. On top of the application service layer there is another layer of components, known as the analytics layer, which has three components: the power manager, the machine learning module, and platform services. The topmost layer is the application layer, which consists of several application platforms. So one can get a clear idea about the technologies involved in their integrated fog, cloud and IoT paradigm. We also found that the virtualization technology plays an important role in their system resources. By following their work, we can easily identify the attributes, features and behaviors of their system resources, what helps contextualizing their work with respect to our interest.

\subsubsection{Syed et al.}
In \cite{18}, the authors present a pattern for the fog computing paradigm. They define pattern as \textit{nothing but a solution to a recurring problem, which is generating in a specific context}. They propose the Pattern-Oriented Software Architecture (POSA) template to describe the pattern of fog computing for a particular context. The pattern illustrates its architecture, including computing, storage and networking services. They understand fog computing as a virtualized platform, which resides on the IoT devices and cloud data centers. According to them, a large number of nodes allows fog computing to collect data efficiently from a distributed platform, while providing low-latency services and supporting the mobility of the fog nodes. They also support location-awareness, heterogeneity, transparency, and big data analytics to the edge, as well as scalability of the platform. Fog computing also endorses the multi-tenancy and multiplicity of providers, and enhances security to the edge devices. By presenting the pattern-oriented software architecture for the fog paradigm, the authors did not explicitly address the classification and taxonomy of their system resources, but following up their work and software architecture we conclude that fog resources are composed and also classified by their computing, storage and networking specification. In addition, we can get the knowledge about the attributes, features and behaviors of their fog resources considered.

\subsubsection{Baccarelli et al.}
In this paper \cite{19}, the authors propose a new dimension of fog computing paradigm: the Fog of Everything (FoE). They present a fog computing architecture, which mainly consists of six blocks: \textit{(1)} the IoE layer, consisting of heterogeneous resource-limited devices, \textit{(2)} the wireless access network, that supports Fog-to-Thing (F2T) and Thing-to-Fog (T2F) communication through TCP/IP connections, \textit{(3)} the interconnected fog nodes, acting as a virtualized cluster head, \textit{(4)} the inter-fog backbone, that provides inter-fog connectivity and inter-fog resource pooling, \textit{(5)} the virtualization layer, that allows each thing to augment its limited resources by exploiting the computing capacity of a corresponding virtual clone, and \textit{(6)} the overlay inter-clone virtual network, that allows P2P inter-clone communication by relying on TCP/IP end-to-end transport connections. Additionally, they divide their FoE model into four different layers, named the cloud layer, the overlay layer, the fog layer and the IoE layer. By following up their work and proposed architecture, we found some attributes, features and behaviors of their system resources. We also found that the networking technologies are involved in their computing paradigm, and that virtualization in another basic technologies, which plays an important role to manage their system resources. In addition, energy or power of the system resources is one of the key points to classify their system resources. Though they have not mentioned explicitly the classification and taxonomy of their system resources, but focusing on their system architecture and attributes of their system, it helps us understanding their system resources and classification.

\subsubsection{Mahmud and Buyya}
In this work \cite{15}, the authors argue that fog computing facilitates location-awareness, mobility support, real-time interactions, scalability, and interoperability, which allows this technology to work efficiently regarding service latency, power consumption, network traffic, capital and operational expenses, and content distribution. These are the key features that should be appropriately handled in a fog computing system to get the best performance. They also discuss different kinds of management issues with severe impact on the fog computing paradigm. Following some previous research works, they review different fog node architectures, application programming platforms, mathematical models, and optimization techniques which attain a certain level of service level objectives (SLO). According to them, these SLOs can be classified into various kind of management aspects to fog computing platform, such as latency management, cost management, network management, computation management, application management, data management, and power management. The main contribution of this paper is the taxonomy of the fog computing paradigm, where the authors consider the following parameters: fog nodes configuration, nodal collaboration, resource/service provisioning metrics, service level objectives, applicable networking system and security concern.

\subsubsection{Perera et al.}
According to the authors in \cite{20}, fog computing supports the dynamic discovery of Internet objects, dynamic configuration and device management, multi-protocol support for both, application and communication level, mobility, and data management at the edge of the network. It enables context discovery and awareness as a facility to the users. In addition, data analytics, security, and privacy can be supported by the fog computing at the edge of the network. In fact, fog computing provides mostly all the features of cloud computing at the edge. In this paper, the authors classify the edge devices into several categories according to their computation capabilities, mainly availability of memory, availability of energy and availability of communication. By following up various research works about fog computing, the authors conclude that cloud companion support and data analytics are the most popular features in these fog computing research efforts.

\subsubsection{Dorsemaine et al.}
In \cite{21}, the authors present a taxonomy for the connected object, where a connected object could be \textit{sensor(s) and/or actuator(s) carrying out a specific function and that can communicate with other equipment; it is part of an infrastructure allowing the transport, storage, processing, and access to the generated data by users or other systems}. According to them, a taxonomy of a connected object can be presented in several categories, such as hardware: storage, memory, processor; software: operating system, APIs; security; network: standards, bandwidth, etc. In this paper, they discuss the different functional attributes regarding interaction with the sensors and actuators, mobility and management of the connected objects.

\subsubsection{Yi et al.}
The authors in \cite{22} argue that there are lots of challenges in the cloud computing paradigm, such as an unreliable latency, and the lack of mobility support and location unawareness for the end users. For this reason, they say that the fog computing architecture facilitates the reliable connectivity and low-latency services to the edge of the network, and that makes an automatic choice over the cloud to perform it in the edge of the network. In this paper, they discuss on the resource provisioning and resource management in fog computing paradigm, and describe the computation offloading strategy in fog computing platforms. They are focused on finding out the policies for monitoring the resources as well as the definition of the cost model for accessing the fog resources. In this paper, the authors did not explicitly address any classification or taxonomy of their system resources, but following up their work we can identify the attributes, behaviors and features of their system resources, as well as the security and privacy issues involved in their system resources. As like other authors, the authors argue that virtualization plays an important role to manage the system resources. 

\subsubsection{Varshney and Simmhan}
In \cite{23}, by comparing edge, fog and cloud computing, the authors present a system architecture of fog computing where they identify the resource characteristics in the fog computing paradigm. The authors say that edge and fog layers have the capabilities of physical and application mobility, privacy sensitivity, and an augmentative runtime environment. In this paper, they also briefly discuss on offloading management and data management in the fog realm. And finally, after discussing on fog computing architectures and following various managerial and application related issues, they identify the challenges for the fog computing paradigm. According to them, there are several technical challenges should be addressed and solved soon: programmability and application-specific customization and optimizations of fog resources, predicting user demand, reducing power and networking consumption, security and privacy, and so on. Following up their contributions, we can get a clear idea about the characteristics, attributes and features of their system resources. Though they have not presented a taxonomy or classification of their system resources, focusing on the characteristics and attributes of their system resources helps identifying the classification the system resources according to their characteristics and attributes.


Note that in our literature survey we have not only focused on those papers discussing the resources classification or taxonomy, but also on those research works which concentrate on managerial related issues in the fog computing paradigm. Managing resources and services properly is essential to run a system efficiently. For instance, in \cite{16} the authors briefly discuss some policy-based service orchestration techniques to manage the services in the fog computing paradigm efficiently. Authors in \cite{19,20,23} present a resource management strategy for fog computing systems, wherein \cite{20} authors are focused on devices management in the fog computing paradigm, and in \cite{19,23} they discuss on the resources and network management in fog computing platforms. In this literature review, we have also found that most researchers consider some use-cases to present their work and, as part of this, they also mention some particular types of devices, which are involved in their proposed system \cite{15,16,17,19,20,24}. And authors in \cite{15,20,21} present some taxonomy for fog computing platforms by considering various use-cases and applications. But most interestingly, and importantly, we have not seen any research work where researchers have explicitly presented some general ontology for a complete fog computing platform, though we found a couple of research works where researchers have given some classifications and taxonomy for such platforms and their resources.

\subsection{IoT Platforms}
\label{subsec:iotp}
In this section we survey those research works in the area of the Internet of Things (IoT). We have found several kind of system architectures or managerial issues in the IoT paradigm to solve the different challenges in this platform by considering different application and use-cases. Interestingly, in these scenarios, from the system architectures and their interactions with the system resources, the researchers propose their views, taxonomy, classification and ontology for the IoT resources. Like in \cite{10,25,26,27,28,29,30,31}, where these researchers propose their system architecture for IoT platforms, but also discuss on the various kind of resource model for this paradigm. In addition, to know a system better, it is essential to know the features and behaviors of the system resources, such in \cite{10,14,25,27,28,30,32,33,34,35}. We have also found that, in some other the works \cite{27,29,30,31,32,33,34}, researchers have given their concern to efficiently manage the IoT system resources. Finally, and more interestingly, we have found some research works \cite{14,27,32,36} where researchers propose explicitly some classification or taxonomy by considering different parameters, and some other \cite{25,26,36}, where the authors discuss on the ontology of the IoT paradigm. So, in the next few paragraphs, we present the various research works related to the IoT paradigm resources classification.\\

\subsubsection{Al-Fuqaha et al.}
The authors in \cite{10} propose a new system architecture for the IoT paradigm. According to the authors, their IoT architecture has five layers: the objects layer, or perception layer, represents the physical sensors of the IoT that aim to collect and process data; the object abstraction layer, transfers data from the objects layer to the service management layer through secure channels; the service management layer, or middleware (pairing) layer, pairs a service with its requester based on addresses and names; the application layer, provides the services requested by customers; and the business layer, or management layer, which manages the overall IoT system activities and services. The IoT communication technologies connect various objects to deliver specific smart services; for example, WiFi, Bluetooth, IEEE 802.15.4, Z-wave, and LTE-Advanced are used in IoT for the communication. In summary, communication technologies and protocols involved in their system resources can be easily identified. In addition, they implicitly mention the attributes and features of their system resources by focusing and discussing the challenges and QoS criteria of the considering system and resources.

\subsubsection{Zhang amd Meng}
Authors in \cite{25} propose a multi-dimensional ontology-based IoT resource model. According to them, the resource model is the combination of the layer structure of IoT, the system architecture, and an ontological presentation and dimensional features of the elements. In their research work, they classify the IoT world into three parts: entity, service, and resource. The entity is the owner or user of resources, such as program, protocol, person, or device. The service is defined as an intermediary between the entity and the resource, which contains IOPE (input, output, precondition, and effect) information, global context information, interface, and specific resource set. And resource is the owner of information. They define the IoT resource model as a combination form of the layer structure, the ontology presentation, and dimensional features of IoT resource elements. The most important contribution of this research work is the ontological representation of the IoT resource model. By presenting such ontology, the researchers analyze the multi-dimensional features of IoT resources.

\subsubsection{Wang et al.}
In \cite{26}, the authors present an Ontology-based Resource Description Model (ORDM) where things in IoT refer to any devices and equipment with functions of perception, control, and processing. They propose things to consist primarily of kinds of sensors, actuators, tags, etc. Their ontology-based resource description model considers the following five classes: attribute, state, control, history, and privacy. The description contents for the IoT devices are classified by examining several aspects such as inherent information of IoT devices containing the classification, property and interface information; working state, acquisition, feedback and stored history information; and the authentication and privacy information of the IoT devices. After presenting their ontology model, they consider a use-case of smart office in order to create a unified resource description for all of the IoT resources. Finally, they discuss on different business related functions for IoT devices.

\subsubsection{Gubbi et al.}
According to the authors in \cite{27} an IoT system can be seen from two perspectives: Internet-centric and Thing centric. In their IoT framework, they considered the WSN (the Network of Things) to reside in the bottom layer. In the middle layer there is the cloud computing platform, and on the top of the layer all the applications that have been running and supported by the WSN and cloud computing layers. The authors identify three components in the IoT paradigm which enables the ubiquitous computing: the hardware, made up of sensors, actuators, and embedded communication hardware; the middleware, consisting on on-demand storage and computing tools for data analytics; and the presentation, consisting of easy to understand visualization and interpretation tools, which can be widely accessed on different platforms. Following up this work we can extract an IoT resources model and identify the attributes and behaviors of the IoT resources.

\subsubsection{Zhao et al.}
In this paper \cite{28} the authors address the demand for making a solution of resource matching and selection in the IoT paradigm. They observed that IoT resources mostly have limited computation capabilities and their exposing resources often operate in dynamic environments. Compared to Web services, the IoT resources are less reliable, their logic is much more straightforward, and their outputs usually represent the observation data of feature-of-interests associated with physical objects. They propose a multidimensional semantic resource model. According to this, each dimension constitutes an ontology model, which includes the well-defined hierarchy, descriptions, and restrictions of the corresponding dimension. Based on this resource model, the semantic matching could accurately measure the similarity of each dimension between resources in parallel, where users are able to aggregate the measurement values of different dimensions according to their preferences. Resources in their model can be RFID tag, sensor, and actuator. We can identify the classification of IoT elements for their considering system paradigm.

\subsubsection{Haller et al.}
In \cite{29} the authors present a domain model of IoT which is providing the template and structures, made by the analysis of various use-cases. According to them, the domain of the IoT can be classified into five core components: the augmented entity (a combination of virtual and physical entity), the user, the device (hardware elements to interact with real-world objects), resources (computational elements), and services. A device is a piece of computing hardware and, more specifically, it is the superclass of that kind of hardware which enables the IoT facilities by establishing the connection between the physical world and the digital world. Resources are considered as the software components, which implement some particular type of functionalities, and can be hosted on a device or anywhere in the network. And services are some specific type of functionality which exposes the standard interfaces and that can be invoked by the users. They show how their IoT domain model can be implemented in a system domain. So considering this work, we can understand the resource definition for their proposed IoT domain and, therefore, get a clear idea about the IoT resources classification.

\subsubsection{Sanchez et al.}
The authors in \cite{30} describe an architectural reference model for open real-world IoT experimentation facilities, as defined in the SmartSantander project. Their architecture is a three-tiered structure, consisting of an IoT device tier, an IoT gateway tier, and a server tier. IoT objects are providing sensing services, and they are capturing and producing data continuously. The IoT model needs to work with some higher storage and computing platform, like cloud computing or fog computing, in order to process and manage those data. Their reference model for IoT system architecture supports three main functional features: experiment support, platform management, and service provision. Considering their use-cases and test-bed implementations, and by following various aspects and management issues, they proposed a taxonomy for the IoT resources in their platform, where these resources can be initially classified into three category, named portal server, gateway, and IoT node. Furthermore, they have classified these three categories according to their host capabilities, type of experiment, service type and participation.

\subsubsection{Kantarci and Mouftah}
In our literature survey we found that some researchers have also considered the cloud computing systems to define the system architecture for IoT, such as in \cite{31}. Specifically, in this research paper they consider a cloud-centric IoT system, which can leverage the efficiency of several applications like as smart health-care, smart transportation systems, smart cities, etc. They propose a cloud platform in a cloud-centric IoT architecture to provide storage resources for aggregated sensing data, as well as computing resources for data analytics and data mining during information retrieval and knowledge discovery on sensing data received via IoT objects. They classify the existing methodologies to address the sensing-as-a-service in the cloud-centric IoT paradigm. According to their proposal, they categorize their quests into three categories: the service provider base search, the GPS-less sensing/scheduling, and the service provider recruitment. Then, the authors address the various research challenges of existing methodologies in a cloud-centric IoT paradigm. In this research work, the authors do not explicitly propose any resources classification or taxonomy for their system paradigm, but when considering the sensing technology, they have defined some sort of classification for their system resources.

\subsubsection{Ahmed et al.}
In this research work \cite{14}, the authors discuss on different kind of networking technologies involved in smart environments, and provide a comparison table between different communication technologies. According to them, WiFi, 3G, 4G, and satellite are the key wireless technologies, where WiFi is mainly used in smart homes, smart cities, smart transportation, smart industries, and smart building environments; 3G and 4G are mostly used in smart cities and smart grid environments; and satellites are used in smart transportation, smart cities, and smart grid environments. In this paper, the authors also discuss on various IoT based smart environments. Considering and following some use-cases, they found some parameters to create a taxonomy for the IoT platform. This taxonomy is based on the following parameters: communication enablers (WiFi, 3G, 4G, satellite), network types (WLAN, WPAN, WAN, MAN, WRAN), technologies (sensing, communication, data fusion, emerging computing, information security), wireless standards (IEEE 802.11, IEEE 802.15.1, IEEE802.15.4), objectives (cost reduction, improve utilization, proactive maintenance, minimal user interaction), and characteristics (prediction capabilities, new enhanced services, remote Monitoring, decision-making capabilities). In this paper, the researchers present a taxonomy for the smart environment, which can be used for making some classification and taxonomy for resources of the IoT paradigm.

\subsubsection{Rayes and Salam}
In this work \cite{32}, the authors characterize the IoT sensing device with three components: the sensor, The microcontroller, and the connectivity. They define eleven types of sensor that can be attached to an IoT platform: temperature, pressure, flow, imaging, fluid level, noise, air pollution, proximity and displacement, infrared, moisture and humidity, and speed sensor. Another way of capturing information from “things” is through the use of RFID, a mechanism to capture information pre-embedded into the so-called tag of a thing or an object using radio waves. It consists of two parts: the tag and the reader. And video tracking is another mechanism to identify and monitor “things”, when sensors and RFID tags are not available. It can also be used in conjunction with sensors and/or RFID to provide a more comprehensive solution. They also classified the actuators, as electrical, mechanical linear, hydraulic, pneumatic, and manual actuators.

\subsubsection{Bermudez-Edo et al.}
The authors in \cite{33} describe IoT concepts in three classes: objects, system or resources, and services. They also classify the IoT devices as sensing devices, actuating devices, and tag devices. From our interest, the most important contribution of this paper is the IoT-Lite ontology, created to be used with a common quantity taxonomy, qu-taxo, to describe the units and quantity kind that IoT devices can measure. This taxonomy is represented by individuals in the ontology and is based on other well-known taxonomies such as qu and qudt. Similarly, some other classes, such as object, service, or attribute, can be linked to a vocabulary to choose the terms from a set of individual and existing concepts. Their proposed ontology is based on Semantic Sensor Network Ontology (SSNO), and they claim the IoT-Lite ontology to be a lightweight version of SSNO.

\subsection{Cloud Computing Platforms}
\label{subsec:ccp}
Cloud Computing is an architectural model that leverages on-demand provisioning and management of computing, and which defines the set of good practices for building an elastic and dynamic set of computing resources to be delivered as a service. In this section, we follow those research works done around the cloud computing paradigm. In this survey we have found that there are more researchers focusing on providing some taxonomy and ontology for cloud resources \cite{4,34,35,36,37,38,39,40,41}, cloud resources management \cite{39,42,43}, and cloud services \cite{37}. However, some researchers have also focused on discussing the cloud paradigm architecture and service models, or identifying the challenges for cloud computing. In the following paragraphs we present the various research works around the cloud paradigm.\\

\subsubsection{Weerasiri et al.}
In \cite{34} the authors present a taxonomy for cloud resources management techniques over the life cycle phases, services and processes, which they call the cloud resource orchestration. They define five dimensions to present the taxonomy for cloud resource orchestration techniques, which are: Resources (resource types, resource entity model, resource access method, resource representation notation), Orchestration Capabilities (primitive actions, orchestration strategies, language paradigm, cross-cutting concern), User Layer (DevOps, application developers, domain experts), Runtime Environment (virtualization technique, execution model, target environment), and Knowledge Reuse (reused artifact, reuse technique). In addition, the cloud providers enable the virtualization techniques through the three categories of cloud resources: infrastructure, platform, and software (as-a-Service). They propose the resource entity to be considered either as an elementary model, where resources do not confide on any other cloud resources, or as a composite model, where cloud resources bring together other elementary resources and also the composite resources to make a significant form of cloud resources. In a cloud computing paradigm, the virtualization technology plays a pivotal role to manage the system and its resources.

\subsubsection{Youseff et al.}
In this paper \cite{35}, the authors present an ontology for the cloud computing paradigm. This has been classified into five layers, such as Applications, Software Environments, Software Infrastructure, Software Kernel, and Hardware. The authors also propose that services offered in the cloud software infrastructure layer could be classified into three part, which are computational resources, data storage, and data communications. The authors also propose the authors propose a pricing model for the cloud computing paradigm, which can be considered into three forms: tiered pricing, per-unit pricing, or subscription-based pricing. In this work, the authors also discussed the various challenges related to the cloud computing paradigm. As per our survey objective, presenting the cloud ontology is the most important contribution of this research work.

\subsubsection{Pittaras et al.}
The authors in \cite{36} propose a distributed resource discovery scheme for multi-cloud infrastructure. They argue that various cloud service providers have their own model to describe cloud resources so, enabling the peers of this multi-cloud architecture requires coming up with a common ontology. They propose the Infrastructure and Network Description Language (INDL) in order to describe the network and infrastructure of a multi-cloud architecture. According to this model, resources are presented as a node (VirtualNode), node components (processor component, memory component, storage component, and switching component), and network element (link, interface). Their main motivation for this research work is allowing a large number of different resource providers, where each peer consists of a resource provider and a resource information service. They also propose their solution to discover the resources among the distributed and heterogeneous resource providers, and this solution also handles dynamic resources status updates and supports infrastructure abstraction policies through use of the semantic web technology. Following this work we can identify the cloud resource components and role of the resource owner, so that might help us to fulfill our interests.

\subsubsection{Zhang et al.}
In \cite{37} the authors present an OWL-based cloud ontology (CoCoOn) which describes the functional and non-functional concepts, attributes and relations of infrastructure services in the cloud computing paradigm. According to them, cloud resources can be classified as Infrastructure-as-a-Service (IaaS), Platform-as-a-Service (PaaS), and Software-as-a-Service (SaaS). They have further classified in their proposed ontology the IaaS resource layer into compute (virtualization), network, and storage (network storage). The authors also propose a system model for CloudRecommender, which can estimate costs and compute saving costs across multiple cloud service providers. In this work, the authors have defined the classification of their system resources, which is of great interest in this survey.

\subsubsection{Arianyan et al.}
In this paper \cite{4}, the authors propose a taxonomy for the cloud computing paradigm. They describe a model to evaluate the cloud products and calculate some value to make some ranking of the products. Their proposed taxonomy model is named the Hierarchical Cloud Taxonomy Engine (HCTE), where they categorize the cloud products into eight major groups, which are deployment model (implementation model, platform model, connectivity), technical features (platform, virtualization, software, hardware support, resource location, standard, billing system, limitation, green computing support), security features (security level, data safety, data removal, incident notification, security monitoring, security in virtual layer, access control, access report), services features (service types, service mobility area, service compatibility, service dependability, user friendliness, service group model, user category, service portability, QoS features), management features (architecture, management support, policies enforcement, SLA monitoring, service orchestration and automation, events management and user report), financial features (charging model, price policy support, payment model, service price evaluation tools), backup and recovery model (data backup, system backup, data after termination, termination alarm, automatic recovery system), and legal aspects, guarantee and support (license type, support policy, support agility, guarantee type, update, language, law enforcement and regularity compliance, popularity). Here the researchers have further classified the deployment models of cloud computing considering the implementation (Private, Public, Hybrid), the platform (distributed or single), and connectivity (full mesh, star, hierarchical). Based on their proposed taxonomy, they present some equations to calculate the score for cloud products so that they can make some cloud product ranking table.

\subsubsection{Fernando et al.}
Researchers in \cite{38} present a new type of cloud computing paradigm, which is mobile in nature. They provide an extensive survey of mobile cloud computing platforms considering various driving issues for the necessity of the mobile cloud computing paradigm. In addition, they present a taxonomy for mobile cloud computing platforms based on the following items: operational issues (offloading method, cost-benefit analysis, mobility management, connection protocols), end user-level issues (incentives, presentation, usability), service and application level issues (application, performance, cloud APIs), privacy, security and trust (general cloud security, mobile cloud security, privacy), context-awareness (service provisioning, risk assessment, resource and common goal identification, energy awareness), and data management (personal data storage on mobile cloud, data access issues, data portability and interoperability, embedded mobile databases). Based on the proposed classification the authors identify the challenges of the mobile cloud computing platform. Though the researchers have not explicitly mentioned the classification or taxonomy for their system resources, this can be extracted from their contributions.

\subsubsection{Singh and Chana}
In this paper \cite{39}, the authors present an extensive survey on resource scheduling and management in the cloud paradigm. They have also classified the cloud resources into two main categories, software resources (application, component service), and hardware resources (processor, network, storage). Resource management in cloud includes two stages, cloud resource provisioning and properly schedule the cloud resources. In this paper, the authors have also focused on the various mapping technology, a different way of resource execution and monitoring strategy. Following the different surveyed research works, they identify lots of challenges in cloud resource scheduling and, also, they discuss the benefits of the cloud resource scheduling techniques. They provide a clear resource classification and, also, we can extract additional resource behaviors from their work.

\subsubsection{Jennings and Stadler}
The authors in \cite{40} revisit several research works in cloud computing and classify the cloud resources into different types. They propose four types, such as compute resources, storage resources, networking resources, and power resources. They also divide the cloud computing paradigm into two models, the service model and the deployment model. The service model can be classified as Infrastructure-as-a-Service, Platform-as-a-Service, and Software-as-a-Service, and the cloud deployment model can be classified as private, public, hybrid, and community. In this paper, the authors provide an extensive survey of resource management strategies, and identify different research challenges for managing the resources in the cloud paradigm. From their contribution, not only we recognize the classification of their system resources, but we can also identify the behaviors and features of their system resources.

\subsubsection{Moscato et al.}
In this paper \cite{41}, the authors define an ontology by identifying the various challenges and problems in the multi-cloud federation structure. The fundamental classes of their ontology are protocol, layer, service models, cloud system visibility, essential characteristics, component, technology, deployment models, framework, actor, property, common state, language, SLA, and abstraction. In their ontology, the resource class is the most complex and, following the Open Cloud Computing Interface (OCCI) specifications, they assume that everything in the cloud system is a cloud resource. In addition, by quoting the National Institute of Standards and Technology (NIST), they classify the cloud computing architecture in five essential characteristics (on-demand self-service, broad network access, resource pooling, rapid elasticity, and measured service), three service models (IaaS, PaaS, SaaS) and four deployment models (public, private, community, and hybrid cloud). They also define the cloud roles at different levels, as service provider, consumer, broker, auditors, and carriers. And finally, by identifying the necessity of cloud federation among the different cloud service providers, the authors propose a federated cloud computing architecture, which they call the mOSAIC Architecture.

\subsubsection{Parikh et al.}
In this research work \cite{42}, the authors classify the cloud resources based on the following utilities: fast computation utility (processor, memory, algorithms, operating system, APIs), storage utility (hard drive, flash drive, database softwares, database servers), communication utility (physical, logical), power / energy utility (cooling devices, UPS), and security utility (trust, authentication, integrity, privacy, availability). Based on this utility based classification, they propose the cloud resource management taxonomy and make some comparison among several cloud resource management algorithms. According to the authors, the cloud computing paradigm has been composed of three kinds of service models (SaaS, PaaS, IaaS) and four types of deployment models (private cloud, public or hosted cloud, community cloud, and hybrid clouds).

\subsubsection{Mustafa et al.}
In \cite{43}, the authors extensively survey several research works about resource management strategies for the cloud computing paradigm. The authors present the service and deployment models, which can be classified as IaaS, PaaS and SaaS for the service model, and private, public, and hybrid for the deployment model. They also propose a taxonomy for cloud resource management strategies, classifying them as energy-aware, SLA-aware, market-oriented, load balanced, network load-aware, resource management for hybrid cloud computing platform, and resource management on mobile cloud platform. The authors also discuss on each classification of their proposed taxonomy, addressing the different strategies for managing the cloud resources. Though the researchers have not explicitly mentioned the classification, taxonomy and ontology for their system resources, we can implicitly obtain the parameters and aspects for classifying their resources.

\subsubsection{Botta et al.}
The authors in \cite{44}, provide an extensive literature survey for the cloudIoT platform by following up two different predominant and popular technologies: cloud computing and IoT. They identify the driving forces for the integration of both technologies by considering several use-cases and applications, and present an architecture for a cloud-IoT platform. The authors also present the service model as IaaS, PaaS and Saas, and the cloud deployment model as private, public, hybrid, and community. After following up some use-cases and applications, they identify several challenges in the cloud-IoT paradigm regarding big data, security, intelligence, network communication, pricing, and SLA enforcement, among others. From this paper we can implicitly extract some attributes, behaviors and features of their system resources, and determine the parameters for classifying their system resources.

As a summary of this category, we have found a couple of research works focused to identifying the characteristics of the cloud computing paradigm \cite{44,45}. We have also found, in \cite{35}, that the researchers have presented some basic concepts for the definition of a pricing model for cloud. There are lots of research works which identify the challenges and issues for the cloud computing platform \cite{38,39,40,41,42,43} and, most importantly, we have found that lots of work have been focused to manage the cloud resources and services \cite{34,37,39,40,42,45}. In this survey, we observed that most research works have been oriented to presenting the cloud computing system architecture and the cloud deployment and service models.

\subsection{Platforms of Related Technologies}
\label{subsec:ocp}
In the previous subsections, we reviewed those research works in the areas of fog computing, IoT, and cloud computing. However, there are still several other works not directly related to those areas, but close to them, which could somehow enhance the findings obtained from our main objectives. So in this section we are going to focus on some research work focused on the following platforms: grid computing, wireless sensor networks (WSN), and mobile edge computing, among others. As our primary intention is to identify the classification, taxonomy and ontology of system resources of various computing paradigms, in this literature survey we have put our focus on those works where researchers have given more concern to identify the resource classification and propose some taxonomy, and following on the architectural and managerial issues of those systems.\\

\subsubsection{Karaoglanoglou and Karatza}
In \cite{46}, the authors identify and discover the appropriate resources in a grid system. They argue that resource discovery in a large-scale grid system is a challenging job, so designing a proper resource catalog would help finding the adequate resources as per the users request. Grid resources are heterogeneous and shared among different organizations, and these resources range from networks, clusters of computers, memory space or storage capacity, computational power (CPU time or CPU cycles), data repositories, attached peripheral devices, sensors, software applications, online instruments, and data. In their grid system, they categorize resources into several levels through a simple matchmaking framework, ranging from limited to powerful capabilities. By considering the memory and storage capacities of grid resources, they define a simple category-table. Then they evaluate the performance of their resource discovery scheme, by considering three parameters: the available disk size, memory space, and the distance in hops a request must travel until it gets satisfied.

\subsubsection{Kaur}
The author in \cite{47} presents a semantic-based resource discovery mechanism for grid environments, considering the grid resource information and usage policies of grid resources. According to the author, \textit{grid is a platform, which consists of a significant amount of distributed and loosely-coupled resources over geographically dispersed locations that equipage together to satisfy the user job/application requirements}. The spread resources in grid form the Virtual Organization (VO), which represents the standard rules to enable sharing and selection. After following different related work, in this paper the author presents a system architecture for grid platform by focusing on the semantic resource discovery scheme, policy matchmaking system, and also the policy aggregator of the grid system. The author also presents an ontology to describe the grid resources. The grid resource information properties consists of the information about operating system type, kernel version, processor speed, number of nodes, RAM speed, hard disk space, storage space, information about the applications, information about the policies, information about the various software, and information about the network.

\subsubsection{Qadir et al.}
In this paper \cite{48}, the authors present a collective overview of the programmable wireless network platform. By following up different application scenarios they presented some challenges in wireless programmable networking system and proposed some new ideas and directions to build the new programmable wireless networking platform. They identified some common trends in programmable wireless networking platforms, and classified some trends for the platform architecture specification, as follows: trends in software-defined wireless networking platform, trends in cognitive wireless networking platform, trends in virtualizable wireless networking platform, and trends in cloud-based wireless networking platform. By analyzing this work we can identify most wireless technologies involved in their system and understand the features, attributes and behaviors of their system resources.

\subsubsection{Oteafy and Hassanein}
Authors in \cite{49} present a novel paradigm in WSN in order to utilize the network resources efficiently. They argue that their proposed architecture is cross-platform, and different applications can use the resources of their proposed WSN platform. Their proposed paradigm is presented in three phases, the resource abstraction and representation, the application representation as a finite set of functional requirements over these resources, and an optimal mapping approach to assigning applications (their functionalities) to the available resources across existing WSNs. In order to identify the primary relations between the IoT and the WSN paradigms, they discuss the various heterogeneous connectivity issues in IoT platform, and also the various dynamic applications working on the WSN platforms. They say that they \textit{introduce an abstraction of all components in a WSN, including ones with confined temporal properties (i.e. resources “passing by”), and extend the definition to encompass IoT components that add to its resource pool (e.g. cell phones, municipal antennas, objects with different access networks, etc.)}. In their paper, the authors considered the sensor nodes and IoT objects as resources, and then they present the core attributes of the resources.

\subsubsection{Ahmed and Ahmed}
In \cite{50}, the authors discuss a particular type of distributed platform, called mobile edge computing (MEC), where computation and processing is usually done in the edge devices. In this paper, the authors identify the taxonomy of mobile edge computing by considering the following parameters: characteristics (proximity, dense geographical distribution, low latency, location awareness, network context information), actors (application developers, content provider, mobile subscribers, OTT players, MEC service provider, software vendors, network equipment vendors), access technologies (Wi-Fi, 3G, 4G, 5G, Wi-Max, Bluetooth), applications (computational offloading, collaborative computing, web content optimization, memory replication, content delivery), objectives (minimize latency, minimize energy, minimize cost, maximize throughput, minimize radio utilization, optimize computational resource), computational platforms (cloud, MEC server, mobile node), and key enablers (cloud and virtualization, high volume servers, network technologies, software development kits, portable devices). They also focused on the pricing model, the scalability, and security. After presenting the taxonomy of the MEC platform, the authors conclude that there are lots of research opportunity to standardize the communication protocol in the MEC paradigm.

\subsubsection{Lee and Hughes}
In this paper \cite{51}, the authors present a new direction of WSN platforms, which strives towards a shared architecture and multi-application paradigm. They introduce a new vision for sensor network paradigm, named tangible cloud, and present the reference architecture and pricing model for this paradigm. They argue the use of an open interface enables resources of the tangible cloud to work with the existing cloud tools, and that makes the tangible cloud resources more inter-operable with the WSN. Following up some use-cases, they propose a reference architecture for their platform, considering the Tangible Infrastructure as a Service, Tangible Platform as a Service, and Tangible Software as a Service. They also introduce the pricing model for their tangible cloud reference architecture. From their contributions we can extract the parameters for classifying their system resources.

\subsubsection{Sharifi et al.}
In \cite{52}, the authors identify the taxonomy of cyber foraging of mobile devices, where computations are done in a powerful non-mobile computation device close to the mobile device. In order to successfully offload a program from a mobile device to a surrogate device, some key metrics should be considered, such as mobile and surrogate specifications, application specifications, network specifications, and context specifications. They also present the taxonomy of cyber foraging approaches by considering the following parameters: offloading type (static, dynamic), surrogate type (static computer, mobile device, cloud), offloading scale (single surrogate, multiple surrogate), live migration, solver location (surrogate, mobile device), code availability (pre-installed RPC, virtual machine, code offloading), offloading granularity (fine-grain application partitioning, coarse-grain whole application migration), parameter of decision (energy, memory, storage, CPU, latency, I/O, responsiveness), and data availability (already available, data offloading from mobile device, transfer from previous surrogate or Internet).


\section{Analysis and Discussion}
\label{sec:ana}
In this section we present a table that summarizes the classification features addressed in the aforementioned 45 research projects and, then, after analyzing the global picture, we present and discuss our findings.

\subsection{Summary of Contributions}
\label{sec:soc}
In our literature survey we have observed that each researcher has different interests and objectives when defining, or suggesting, their resources classification or taxonomy. Some of them from a system architecture perspective and according to their platform scope (cloud, fog, IoT, etc.), some others from a resources and services management perspective, or some others concerned about understanding the system behaviors and features, and addressing the various challenges of those systems. So we have identified the parameters suggested in all surveyed research works, organized, and assessed them according to the following categories:
\begin{itemize}
\item Hardware: Processor, Memory, Storage, Power, GPU, and FPGA
\item Software: Operating System, Application and APIs, and Database Product Information
\item Network: Bandwidth, Technology and Standards, Applicable Networking System, Communicating Techniques, and Deployment Model
\item Security: Authenticity, Intrusion, Privacy, Trust, Encryption, Denial of Service Attack, Identification, Accountability, Confidentiality, and Authorization
\item Data: Source and Owner, Type and Size, Content Format, Analysis Type, Processing Framework or Infrastructure
\item IoT: Type of Sensors, Actuators, and RFID tags
\item Cost: Computation, Communication, and Deployment
\item Features and Behaviors: Scalability, Fault-Tolerance, Location-Awareness, Availability, Agility / Real-Time, Autonomy, Mobility, Multi-Tenancy, Reliability, Context Information, Programmability and Virtualization, Ubiquitous, Heterogeneity, On-Demand, Geo-Distribution, Flexibility / Interoperability, Transparency / Openness, and Proximity
\item Management: Latency, Power, Data, Network, Cost, Application, Computation, Operational, Service, Policies, User-Actor, and SLA
\item Use-Cases or Applicable System: Type of Use-Cases
\end{itemize}
Then, considering the upper mentioned parameters, we have built the Table 1, where all the fields for these categories (rows in the table) have been filled up according to the description of each research work (columns in the table). The box corresponding to each category which is addressed in the corresponding research work has been shadowed, so providing a graphical snapshot of what categories are more relevant for the different research areas and, therefore, allowing the extraction of many valuable findings, as presented in the next subsection.

\begin{figure*}
\centering
\includegraphics[width=\textwidth,height=\textheight]{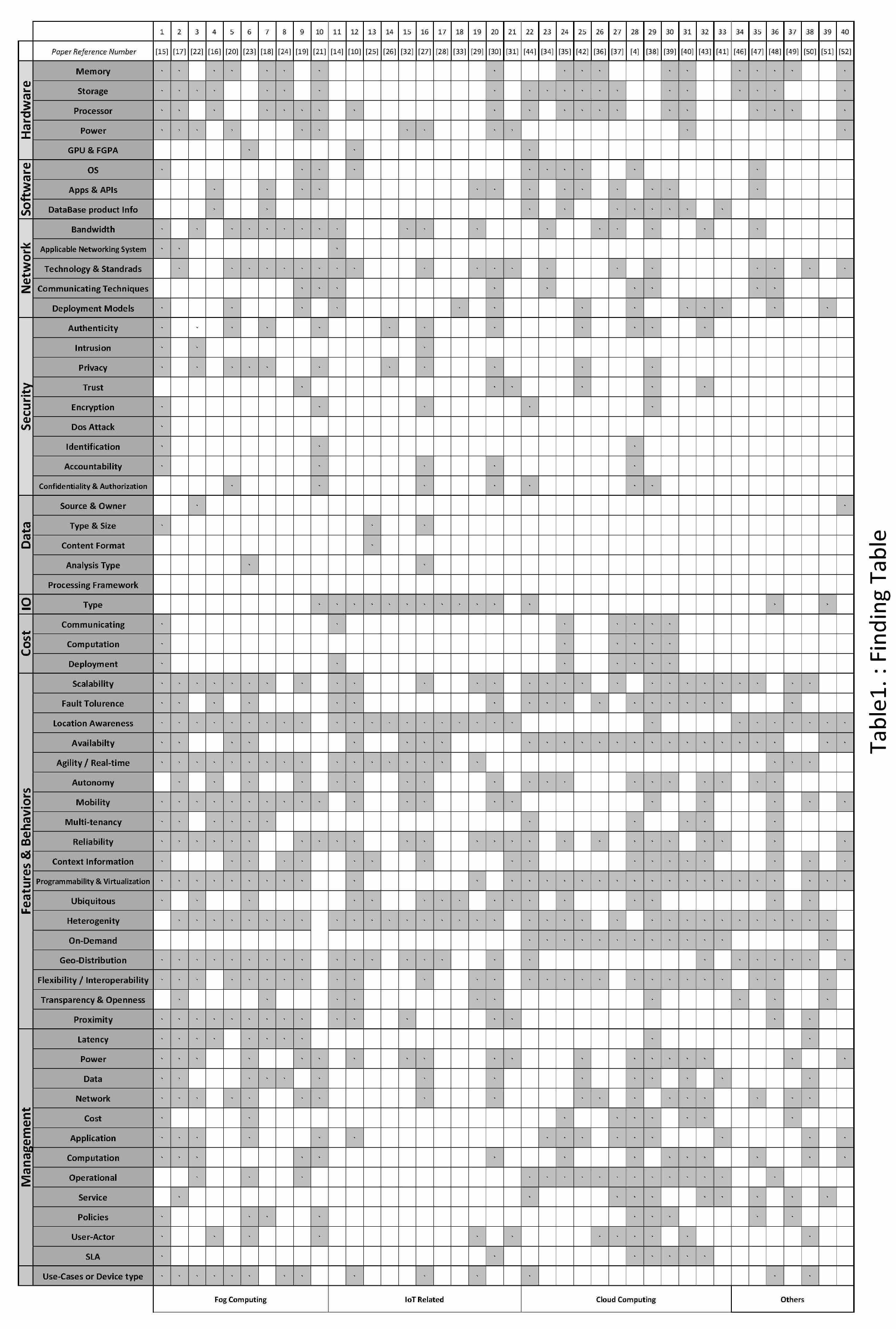}
\label{lab:image}
\end{figure*}

\subsection{Discussion of Contributions}
\label{sec:synop}
After analyzing the table, different synoptic findings can be extracted. We have organized these findings following the same categories used to organize the table, which are: Hardware related findings, Software related findings, Networking related findings, Security related findings, Data related findings, IoT devices related findings, Cost related findings, Features related findings, and Management related findings. Next, we present each of these findings.

\textbf{Hardware related findings}. We have observed that most researchers in the fog computing and IoT based research works are concerned about an efficient power management. This has sense, since many small devices at the edge (for instance, smartphones) as well as many sensors, are battery powered. In addition, except for the IoT based research works, almost every other project focuses on the hardware components (i.e., processor, memory, storage). Interestingly, we have only found three papers \cite{10,23,44} where researchers consider other hardware components, such as GPU and FGPA, to represent the characteristics of their considered computing platform.

\textbf{Software related findings}. With respect to the software related issues, we have found that software is not important in most IoT based research. Exceptionally, only in a couple of these works \cite{29,30}, the authors have considered the APIs involved in their system resources in order to create a catalog of services. Conversely, this attributes category gains relevancy in the cloud related systems, where the software to be classified and organized becomes more diverse. It is also relevant for some fog computing-based research works \cite{16,18,19,21} where researchers are concerned about identifying the various heterogeneous OS, apps, and APIs involved with the fog and edge devices.

\textbf{Networking related findings}. The network related attributes are interesting, mainly, for fog and edge computing research works, with main concern about network bandwidth management. As the variety in fog and IoT devices is great, many researchers are focused on identifying the standards and protocols for these paradigms. We also found that in some papers, such as \cite{14,15,17}, they discuss various types of networking devices (switches, routers, set-top boxes) in order to present their system architecture. But most interestingly, we have observed that some cloud-based research works are also interested in network bandwidth in order to manage cloud resources \cite{34,43,45}, and discovering resources \cite{36} or services\cite{37}.

\textbf{Security related findings}. The most relevant security attributes addressed are authenticity, privacy, trustiness, and confidentiality. In addition, with respect to the security category, most research works are either concerned in several security aspects, or in none. It is also worth noticing that the DOS attack is only considered as an eventual security challenge in one research paper from the fog computing paradigm \cite{15}. In any case, from this literature survey, security is one of the least addresses set of attributes.

\textbf{Data related findings}. Though data is one of the key components of any computing paradigm, we found that this attribute is one of the least considered. Only in some IoT related research works has been partially considered, addressing concepts such as data source, size, and content type information. One exception is \cite{52} where researchers identify the taxonomy of cyber foraging at mobile devices. According to them, data availability is one of the critical aspects of the cyber foraging technique, and that helps classifying mobile devices in this category.

\textbf{IoT related findings}. This category is expected in most IoT-based research works, as expected. Researchers have discussed on the type of the sensors, actuators, and RFID tags. We have also observed in \cite{21,44} that, to address the connected objects and the cloud-IoT paradigm, researchers have focused to identifying the type of sensors, actuators, and RFID in their systems. In fact, these two papers focus on integrated fog and cloud platforms, so this fact is actually quite natural.

\textbf{Cost related findings}. This category is certainly the one with least interest. It is being considered mainly in cloud based research works. In fact, it is evident that cost is a crucial factor that should be considered before presenting some new proposal for making the cloud much more efficient.

\textbf{Features related findings}. The features and behaviors category includes a variable set of specific attributes. We observe that the cloud computing paradigm has some unique features, such as fault-tolerance, centralization, on-demand, etc; but no concern for transparency and openness, nor geographical-related issues. In other computing platforms (i.e., fog, IoT, grid, and WSN) resources are geographically distributed and near to the edge of the network, so they provide real-time services, and also these systems are very much aware of their working location. We also found that virtualization is mostly related to the fog and cloud computing paradigms.

\textbf{Management related findings}. Management is one category mainly addressed in the areas of fog computing and cloud computing. Within this category, the cloud paradigm seems more interested in features such as cost, operational, and SLA, whereas the fog paradigm is more interested in latency, power management, and data management. Interestingly, in the IoT area there is minimal concern about these attributes.


\section{Conclusion}
\label{sec:conc}
In this paper we have developed an exhaustive survey about the different research works which addresses, somehow, any sort of resources categorization, classification, taxonomy or ontology, in the context of smart environments technology. Although only some authors explicitly use any of these terms in their works, some others propose implicitly some resources classification as part of their system architecture organization or management. We have tried to include all of them in this literature survey. With this survey we pretend to find out the technological challenges and gaps among the different computing paradigms and technologies.

We have observed that each researcher has different interests and objectives when defining, or suggesting, their resources classification or taxonomy. Some of them from a system architecture perspective and according to their platform scope, some others from a resources and services management perspective, or some others concerned about understanding the system behaviors and features. As part of this research work, we have designed and presented a table that summarizes all surveyed works and provides a compact and graphical snapshot of the current classification trends.

The goal and primary motivation of this survey when identifying the resource classification has been to understand the current state of the art and identify the gaps between the different computing paradigms involved in smart environment scenarios. As a result of our survey, we have found that it is essential to consider together several computing paradigms and technologies, and that there is not, yet, any research work that integrates a merged resources classification, taxonomy or ontology required in such heterogeneous scenarios. With these goals as a starting point, we will extend our research with the definition and assessment of a new and comprehensive resources ontology that considers all relevant technological paradigms associated to the smart environments, including fog computing, cloud computing, and the IoT technologies.

\section*{Acknowledgment}
This work was supported by the Spanish Ministry of Economy and Competitiveness
and the European Regional Development Fund, under contract TEC2015-66220-R
(MINECO/FEDER), and by the H2020 EU mF2C project reference 730929.

\ifCLASSOPTIONcaptionsoff
  \newpage
\fi

\end{document}